\documentclass[12pt]{article}
\usepackage[dvips]{graphicx}
\usepackage[latin1]{inputenc}
\usepackage{amsmath}
\usepackage{amsfonts}
\usepackage{amssymb}
\newcommand{\be}{\begin{equation}}
\newcommand{\ee}{\end{equation}}
\newcommand{\ben}{\begin{eqnarray}}
\newcommand{\een}{\end{eqnarray}}
\newcommand{\non}{\nonumber}
\newcommand{\foot}{\footnote}

\begin{document}

\title{\bf\Large High-energy neutrino emission from X-ray binaries}

\author{ {\normalsize\bf Hugo R. Christiansen$^a$\thanks{Electronic addresses: hugo@uece.br},
 {\normalsize\bf Mariana Orellana$^b$\thanks{morellana@iar.unlp.edu.ar}},
 {\normalsize and} {\normalsize\bf Gustavo E. Romero$^b$\thanks{romero@iar.unlp.edu.ar}} }\\
 \\
{\normalsize\it $^a$State Univesity of Cear\'a, Physics Dept., Av. Paranjana 1700,
60740-000}\\ {\normalsize\it Fortaleza - CE, Brazil}\\
{\normalsize\it $^b$Instituto Argentino de Radioastronom\'{\i}a (IAR), C.C. 5,\ 1894}\\
{\normalsize\it Villa Elisa - Bs.As., Argentina}}

\date{}
\maketitle
\begin{abstract}

\noindent We show that high-energy neutrinos can be efficiently
produced in X-ray binaries with relativistic jets and high-mass
primary stars. We consider a system where the star presents a dense
equatorial wind and the jet has a small content of relativistic
protons. In this scenario, neutrinos and correlated gamma-rays
result from $pp$ interactions and the subsequent pion decays. As a
particular example we consider the microquasar  LS I $+61 \;303$.
Above 1 TeV, we obtain a mean-orbital $\nu_{\mu}$-luminosity of
$\sim 5 \; 10^{34}$ erg/s which can be related to an event rate of
4-5 muon-type neutrinos per kilometer-squared per year after
considering the signal attenuation due to maximal neutrino
oscillations. The maximal neutrino energies here considered will
range between 20 and 85 TeV along the orbit. The local infrared
photon-field is responsible for opacity effects on the associated
gamma radiation at high energies, but below 50 GeV the source could
be detected by MAGIC telescope. GLAST observations at
$E_{\gamma}>100$ MeV should also reveal a strong source.

\end{abstract}

 \vskip 1cm

{\tt To Appear in Phys.~Rev.~D (April 2006 issue)\foot{Submitted on
12th August 2005.}.}

\section{Introduction}

The study of high-energy neutrinos from galactic sources is expected
to provide important clues for the understanding of the origin of
cosmic rays in our Galaxy. Astrophysical sources of high-energy
neutrinos should have relativistic hadrons and suitable targets for
them, such as radiation and/or matter fields. Rotating magnetized
neutron stars are well-known particle accelerators. When they make
part of a binary system, relativistic particles may find convenient
targets in the companion star and high-energy interactions can take
place. It is then natural to consider accreting neutron stars or,
more generally, X-ray binaries, as potential neutrino sources (see
Bednarek et al. 2005 and references therein\cite{bednaetal}).

Neutron stars with high magnetic fields can disrupt the accreting flow, which
is channeled through the field lines to the magnetic poles of the system. The
magnetosphere of such systems presents electrostatic gaps where protons can be
accelerated up to very high energies. Actually, even in the absence of
relativistic jets, it is possible to attain hundreds of TeVs  (Cheng \&
Ruderman 1989, 1991 \cite{cheng89,cheng91}). The accelerated protons, which
move along the closed field lines, can impact on the accreting material
producing gamma-rays (Cheng et al. 1992a \cite{cheng92a}, Romero et al. 2001
\cite{rom01}, Orellana \& Romero 2005 \cite{ore05}) and neutrinos (Cheng et
al. 1992b \cite{cheng92b}, Anchordoqui et al. 2003 \cite{luis03}).

If the magnetic field of the neutron star is not very high, then
accretion/ejection phenomena can appear, and the X-ray binary can display
relativistic jets, as in the well-known cases of Sco X-1 (Fender et al. 2004
\cite{fender04}) and LS I $+61 \;303$ (Massi et al. 2001, 2004
\cite{massi01,massi04}). The X-ray binary is then called a 'microquasar'. In
this case, a fraction of the jet hadrons can reach much higher energies,
up to a hundred of PeVs or more, depending on the parameters of the system
(see below).

Microquasars can be powered either by a weakly magnetized neutron star or by a
black hole (as, for instance, Cygnus X-1). Some of them are suspected to be
gamma-ray sources (Paredes et al. 2000 \cite{paredes00}, Kaufman-Bernad\'o et
al. 2002 \cite{kf02}, Bosch-Ramon et al. 2005a \cite{br05a}). Recently,
Aharonian et al. (2005) \cite{aha05} have detected very high-energy emission
from the microquasar LS 5039 using the High Energy Stereoscopic System (HESS).
In microquasars, the presence of relativistic hadrons in the jets can lead to
neutrino production through photo-hadron (Levinson \& Waxman 2001 \cite{lw01},
Distefano et al. 2002 \cite{distefano02}) or proton-proton (Romero et al. 2003
\cite{rom03}, Romero \& Orellana 2005 \cite{rom05}, Torres et al. 2005
\cite{torres05}, Bednarek 2005 \cite{bedna05}) interactions. In the latter
case, the target protons are provided by the stellar wind of the companion.

In the case studied by Romero et al. (2003) \cite{rom03}, which was the base of
subsequent models, the wind is assumed to be spherically symmetric. This,
however, is not always the case. In particular, microquasars with Be stellar
companions present slow and dense equatorial winds that form a circumstellar
disk around the primary star and the compact object moves inside. In the
present paper we will study how the interaction of the jet with this material
can lead to a prominent neutrino source. We shall focus on a specific object,
LS I $+61 \;303$, for which all basic parameters are rather well-determined in
order to make quantitative predictions that can be tested with the new
generation of neutrino telescopes (IceCube, in this case \cite{icecube}). Our
results, however, will have general interest for any microquasar with
non-spherically symmetric winds. We shall start with a brief description of
LS I $+61 \;303$.

\section{The microquasar LS I $+61 \;303$}

LS I $+61 \;303$ is a Be/X-ray binary system that presents unusually strong and
variable radio emission (Gregory \& Taylor 1978 \cite{gt78}). The X-ray
emission is weaker than in other objects of the same class (e.g. Greiner \&
Rau 2001 \cite{gr01}) and shows a modulation with the radio period (Paredes et
al. 1997 \cite{paredes97}). The most recent determination of the orbital
parameters (Casares et al. 2005 \cite{casares05}) indicates that the
eccentricity of the system is $0.72\pm0.15$ and that the orbital inclination
is $\sim 30^{\circ}\pm 20^{\circ}$. The best determination of the orbital
period ($P=26.4960\pm0.0028$) comes from radio data (Gregory 2002
\cite{gregory02}). The primary star is a B0 V with a dense equatorial wind.
Its distance is $\sim 2$ kpc. The X-ray/radio outbursts are triggered 2.5-4
days after the periastron passage of the compact object, usually thought to be
a neutron star. These outbursts can last until well beyond the apastron
passage.

Recently, Massi et al. (2001) \cite{massi01} have detected the existence of
relativistic radio jets in LS I $+61 \, 303$. These jets seem to extend up to
about 400 AU from the compact object (Massi et al. 2004 \cite{massietal}).

LS I $+61 \, 303$ has long been associated with a gamma-ray source. First with
the COS-B source CG135+01, and later on with 3EG J0241+6103 (Gregory \& Taylor
1978 \cite{gt78}, Kniffen et al 1997 \cite{kni97}). The gamma-ray emission is
clearly variable (Tavani et al. 1998 \cite{tava98}) and has been recently
shown that the peak of the gamma-ray lightcurve is consistent with the
periastron passage (Massi 2004 \cite{massi04}), contrary to what happens with
the radio/X-ray emission, which peaks {\sl after} the passage. A complete an
updated summary of the source is given by Massi 2005 \cite{massi05}.

The matter content of microquasar jets is unknown, although in the case of SS
433 iron X-ray line observations have proved the presence of ions in the jets
(Kotani et al. 1994, 1996 \cite{kota94, kota96}; Migliari et al. 2002
\cite{migliardi02}). In the present paper we will assume that relativistic
protons are part of the content of the observed jets in LS I $+61 \, 303$. In
the next section we will describe the basic features of the model (see Romero
et al. 2005 \cite{hugo} for a detailed discussion of the gamma-ray emission),
and then we will present the calculations and results.

\section{Model\label{model}}

The system under consideration consists of two objects. The primary is a
B-type star that generates a radially outflowing wind. The other is a compact
object (CO hereafter, probably a neutron star) moving around in a Keplerian
orbit. Their relative position is given  by $r(\psi)=a (1-e^2)/(1-e \cos
(\psi))$, where $\psi$ is the orbital phase, $a$ is the semi-major axis of the
ellipse and $e$ its eccentricity \footnote{Notice that here $\psi$ is {\sl
not} the radio phase, which amounts 0.23 $2\pi$ at the periastron (Casares et
al. 2005) \cite{casares05}}. The primary presents a nearly equatorial
circumstellar disk with a half-opening angle  $\phi=15^\circ$. Its density is
given by $\rho_{\rm w}(r)=\rho_0({r}/{R_*})^{-n}$ and the wind velocity reads
$v_{\rm w} = v_0 (r/R_*)^{n-2}$ by continuity. We will adopt $n=3.2$ following
Mart\'{\i} and Paredes 1995 \cite{paredes95} (see also Gregory \& Neish, 2002
\cite{gn02}). The wind accretion rate onto the compact object is given by

\be \dot{M}_{\rm c}=\frac{4 \pi (G\,M_{\rm c})^2 \rho_w(r)}{v_{\rm rel}^3},
\label{macc} \ee
where $M_{\rm c}$ is the CO mass and $v_{\rm rel}$ is its velocity relative to
the circumstellar wind. The kinetic jet power  $Q_{\rm j}$ is coupled to
$\dot{M}_{\rm c}$  by $ Q_{\rm j}=q_{\rm j} \dot{M}_{\rm c} c^2 $ (Falcke \&
Biermann 1995 \cite{fb95}). In the jet-disk symbiosis model $q_{\rm j}\sim
0.1$ for microquasars in the low-hard state (Fender 2001 \cite{fender01}).
Only a small fraction of the jet particles are highly relativistic hadrons
($\sim 10^{-2}$), and
these are confined by the pressure of the cold particles expanding laterally
at the local sound speed (see Bosch-Ramon et al. 2005b \cite{br05b} for a
detailed discussion). In the present approach, most of the jet power will
consist of cold protons ejected with a macroscopic Lorentz factor $\Gamma\sim
1.25$ (Massi et al. 2001 \cite{massi01}).

\begin{figure*}[t]
\centering
\includegraphics[width=0.45\textwidth,height=6.5cm]{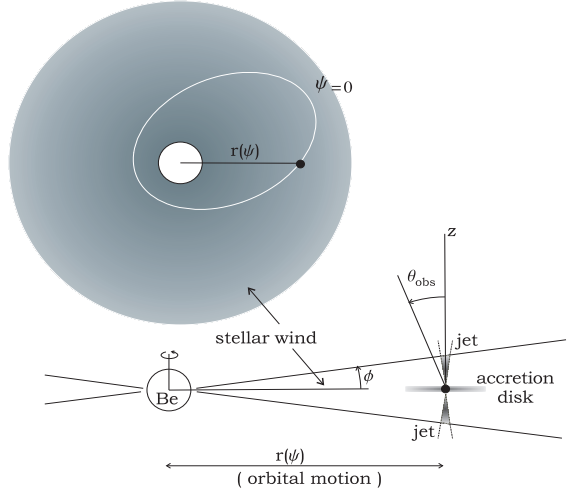}
\caption{A sketch of the binary system with its elements in
 detail.}\label{fig:esquema}
\end{figure*}

The jet will be a cone with a radius $R_{\rm j}(z)= z (R_0/z_0)$, where $z_0=
10^7$ cm is the injection point and $R_0=z_0/10$ is the initial radius of the
jet (see Romero et al. 2003 \cite{rom03} and Bosch-Ramon et al. 2005a
\cite{br05a} for additional details). The jet axis, $z$, will be taken normal
to the orbital plane. In Figure \ref{fig:esquema} we show a sketch of the
general situation.

We will use a power-law relativistic proton spectrum $N'_{p}(E'_{p})= K_p\;
{E'}_{p}^{-\alpha}$ (the prime refers to the jet frame). The corresponding
relativistic proton flux is $J'_{p}( E'_{p})= (c/4\pi) N'_{p}(E'_{p})$
evolving with $z$ as $ J'_p(E'_p)=\frac{c K_0}{4 \pi} \left(z_0/z\right)^2
{E'_p}^{-\alpha}, $ (conservation of the number of particles is assumed, see
Ghisellini et al. 1985 \cite{ghise85}). Using relativistic invariants, it can
be shown that the proton flux in the lab (observer) frame becomes (e.g.
Purmohammad \& Samimi 2001 \cite{ps01})

\begin{equation}
J_p(E_p,\theta)=\frac{c K_0}{4 \pi} \left(\frac{z_0}{z}\right)^
2 \frac{\Gamma^{-\alpha+1} \left(E_p-\beta_{\rm b}
\sqrt{E_p^2-m_p^2c^4} \cos \theta\right)^{-\alpha}}{\left[\sin ^2
\theta + \Gamma^2 \left( \cos \theta - \frac{\beta_{\rm b}
E_p}{\sqrt{E_p^2-m_p^2 c^4}}\right)^2\right]^{1/2}}. \label{Jp_lab}
\end{equation}
The angle subtended by the proton velocity direction and the jet axis will be
roughly the same as that of the emerging photon ($\theta \approx \theta_{\rm
obs}$). $\beta_{\rm b}$ is the bulk velocity in units of $c$, and $\alpha=2.2$
characterizes the power law spectrum (see the list of parameters in Table 1).
We adopt a value of 2.2 for the proton power-law index instead of the
canonical one of 2.0 given by first order diffusive shock acceleration. This
is in order to match the GeV gamma-ray spectrum observed by EGRET
\cite{hartman}. Deviation from the canonical value are likely due to nonlinear
effects (see, e.g., Malkov \& Drury 2001 \cite{drury}). The normalization
constant $K_0$ and the number density ${n_0}'$ of particles flowing in the jet
at $R_0$ can be determined as in Romero et al. (2003) \cite{rom03}.

As long as the particle gyro-radius is smaller than the radius of the jet, the
matter from the wind can penetrate the jet diffusing into it\foot{This imposes
a constraint onto the value of the magnetic field in the jet: $B_{\rm jet}\geq
E^{\rm p,\; wind}_{\rm k}/(e R)$, where $E^{\rm p,\; wind}_k = m_p\,v_{\rm
rel}^2/2$ is the kinetic energy of the cold protons in the slow stellar wind
(for $E^{\rm p,\; wind}_{\rm k}$ maximum, at periastron, results $B_{\rm
jet}\geq 2.8\,10^{-6}$ G). This condition, as we will see, is assured.}.
However, some effects, like shock formation on the boundary layers, could
prevent some particles from entering into the jet. Actually, the problem of
matter exchange through the boundary layers of a relativistic jet is a
difficult one. Contrary to cases studied in the literature, where the external
medium has no velocity {towards} the jet (e.g. Ostrowski 1998
\cite{ostrowski}), here the wind impacts on the jet side with a velocity that
can reach significant values (hundreds of km s$^{-1}$). A detailed study of the
penetration of the wind into the relativistic flow is beyond the scope of the
present paper, so we will treat the problem in a phenomenological way. This
can be done by means of a ``penetration factor'' $f_{\rm p}$ that takes into
account particle rejection from the boundary. We will adopt $f_{\rm p}\sim
0.1$, in order to reproduce the observed gamma-ray flux at GeV energies, where
opacity effects due to pair creation are unimportant.


We will make all calculations in the lab frame, where the cross sections for
proton interactions have suitable parameterizations. Neutral pion decay after
high energy proton collisions is a natural channel for high energy gamma-ray
production. The differential gamma-ray emissivity from $\pi^0$-decays can be
expressed as
 (e.g. Aharonian \& Atoyan 1996 \cite{aa96}):
\begin{equation}
q_{\gamma}(E_{\gamma},\theta)= 4 \pi\,\eta_{\rm A} \sigma_{pp}(E_p)
\frac{2Z^{(\alpha)}_{p\rightarrow\pi^0}}{\alpha}\;J_p(E_{\gamma},\theta) ,
\label{q}
\end{equation}
(in  ph s$^{-1}$ sr$^{-1}$ erg$^{-1}$), where
$Z^{(\alpha)}_{p\rightarrow\pi^0}$ is the so-called spectrum-weighted moment
of the inclusive cross-section and it is related to the fraction of kinetic
proton energy transferred to the pions (see Gaisser 1990 \cite{gaisser90}).
The parameter $\eta_{\rm A}$ takes into account the contribution from
different nuclei in the wind. For standard composition of cosmic rays and
interstellar medium $\eta_{\rm A}\sim 1.4$. The proton flux distribution
$J_p(E_{\gamma})$ (see eq.(\ref{Jp_lab})) is evaluated at $E=E_{\gamma}$, and
$\sigma_{pp}(E_p)\approx 30 \times [0.95 + 0.06 \log \;(E_p/{\rm GeV})]\;\;
(\rm mb)$ \cite{aa96} is the cross section for inelastic $pp$ interactions at
energy $E_p\approx 6 \xi_{\pi^{0}} E_{\gamma}/K$, for $E_{\gamma}\geq 1$ GeV.
Here $K$ is the inelasticity coefficient, and $\xi_{\pi^{0}}$ represents the
pion multiplicity. For $\theta$ we will adopt a viewing angle of $\theta_{\rm
obs} = 30^\circ$ in accordance with the average value given by Casares et al.
(2005) \cite{casares05}.

The spectral gamma ray intensity (in ph s$^{-1}$ erg$^{-1}$) is
\begin{equation}
I_{\gamma}(E_{\gamma},\theta)= \int_V \ d^3\vec{r} \
n(\vec{r})\,q_{\gamma}(E_\gamma,\theta),
 \label{Intensity}
\end{equation}
where $V$ is the interaction volume between the jet and the circumstellar
disk. The particle density of the wind that penetrates the jet is $n(r)\approx
f_{\rm p} \rho_w(r)/m_p$, and the generated luminosity in a given energy band
results

\be L_{\gamma}(E_{\gamma}^{a,b},\theta)= \int_{E_{\gamma a}}^{E_{\gamma b}}\
dE_{\gamma}\ E_\gamma  \ I_{\gamma}(E_{\gamma},\theta) \label{luminosity}\ee

Since we are interested in predicting the total gamma ray luminosity
measurable on Earth, we must evaluate this integral above a threshold of
detection which shall depend on the telescope characteristics. On the other
hand, we should keep in mind that $E_{\gamma}$ cannot exceed the maximum gamma
ray energy available from the hadronic processes described so far. We will
address
 this issue in more detail in the following section.

\section{Hadronic production of neutrinos and gamma rays}


In order to discuss the hadronic origin of high energy gamma-rays and
neutrinos we have to analyze in more detail pion luminosity and decay.
Regarding gamma-rays production, since the branching ratio of charged pions
into photons (plus leptons) is about 6 orders of magnitude smaller than that
of $\pi^0$'s into photons (alone) \cite{PDG} we shall first concentrate on
these neutral parents. Actually, high-energy neutral pions can be produced by
relativistic proton collisions in either $pp$ or $p\gamma$ interactions
depending on the relative target density of photons and protons in the source
region (where the protons are accelerated) and on their relative branching
ratios. On the other hand, charged pions will be responsible for high energy
neutrinos. In any case, note that both photon and neutrino production at the
source are closely related. The $\nu_\mu + \bar\nu_\mu$ differential neutrino
flux ($dN_\nu/dE_\nu$) produced by the decay of charged pions can be actually
derived from the differential $\gamma$-ray flux $(dN_\gamma/dE_\gamma)$.
Following Alvarez-Mu\~niz and Halzen (2002) \cite{amh02}, we will find the
neutrino intensity and spectral flux by means of an identity related to the
conservation of energy (see also Stecker 1979 \cite{stecker79, salamon96})
\begin{equation}
\int_{E_{\gamma}^{\rm min}}^{E_{\gamma}^{\rm max}} E_\gamma {dN_\gamma\over
dE_\gamma} dE_\gamma = D \int_{E_{\nu}^{\rm min}}^{E_{\nu}^{\rm max}} E_\nu
{dN_\nu\over dE_\nu} dE_\nu~. \label{conservation}
\end{equation}
Here ${E_{\gamma}^{\rm min}}$ ($E_{\gamma}^{\rm max}$) is the minimum
(maximum) energy of the photons that have a hadronic origin and ${E_{\nu}^{\rm
min}}$ and ${E_{\nu}^{\rm max}}$ are the corresponding minimum and maximum
energies of the neutrinos.
The relationship between both integrals, as given by $D$, depends on the
energy distribution among the particles resulting from the inelastic
collision. As a consequence, it will depend on whether the $\pi^0$'s are of
$pp$ or $p\gamma$ origin. Its value can be obtained from routine particle
physics calculations and some kinematic assumptions. Let us admit that in $pp$
interactions 1/3 of the proton energy goes into each pion flavor on average.
We can further assume that in the (roughly 99.9$\%$) pion decay chains,
\begin{eqnarray}
\pi^+ \rightarrow \mu^+ \nu_\mu \rightarrow e^+ \nu_e \bar\nu_\mu \nu_\mu
\hskip 1cm \pi^- \rightarrow \mu^- \bar\nu_\mu \rightarrow e^- \bar\nu_e
\nu_\mu \bar\nu_\mu\hskip 1cm \pi^0 \rightarrow \gamma\gamma,\non \een two
muon-neutrinos (and two muon-antineutrinos) are produced in the charged
channel with energy $E_\pi/4$, for every photon with energy $E_\pi/2$ in the
neutral channel. Therefore the energy in neutrinos matches the energy in
photons and $D=1$.

The relevant parameters to relate the neutrino flux to the $\gamma$-ray flux
are the maximum and minimum energies of the produced photons and neutrinos,
appearing as integration limits in Eq.\,(\ref{conservation}). The maximum
neutrino energy is fixed by the maximum energy of the accelerated protons
($E_p^{\rm max}$) which can be conservatively obtained from the maximum
observed $\gamma$-ray energy $E_{\gamma}^{\rm max}$. Following our previous
assumptions
\begin{equation}
E_p^{\rm max}=6 E_{\gamma}^{\rm max}~,~~~E_{\nu}^{\rm max}={1\over 2}
E_{\gamma}^{\rm max}~, \label{enumax}
\end{equation}
for the $pp$ case. The minimum gamma and neutrino energies are fixed by the
threshold for pion production. For the $pp$ case
\begin{equation}
E_p^{\rm min}=\Gamma ~ {(2m_p+m_\pi)^2-2m_p^2 \over 2m_p}, \label{ethpp}
\end{equation}
where $\Gamma$ is the Lorentz factor of the accelerator relative to the
observer. The average minimum neutrino energy is obtained from $E_p^{\rm min}$
using the same relations of Eq. (\ref{enumax}).

In $p\gamma$ interactions we can assume that neutrinos are predominantly
produced via the $\Delta$-resonance. In 1/3 of the interactions a $\pi^+$ is
produced which decays into two neutrinos of energy $E_\pi/4$, and in the other
$2/3$ of the interactions a $\pi^0$ is produced which decays into two photons
of energy $E_\pi/2$. Therefore $D=4$. The minimum proton energy is given by
the threshold for production of pions, and the maximum neutrino energy is
about $5\%$ of the maximum energy to which protons are accelerated, which is
much more than that obtained from $pp$ collisions. However, when the
production of neutrinos in $p\gamma$ collisions occurs in the acceleration
region, the efficiency of conversion into relativistic hadrons is lower than
necessary. This is due to the fact that the threshold for electron-positron
production is about two orders of magnitude below that for pion production
along with the fact that, in this case, the $e^{\pm}$ production cross section
is significantly larger. As a result, most of the energy from the acceleration
mechanism is transferred to leptons, radiating plenty of photons but no
neutrinos.

On the other hand, when purely hadronic collisions are considered, the
situation is reverted and pion production becomes the natural production
channel. Relativistic protons in the jet will interact with target protons in
the wind through the reaction channel $p+p\rightarrow p+p+ \xi_{\pi^{0}}
\pi^{0} + \xi_{\pi^{\pm}} (\pi^{+} + \pi^{-})$. Then pion decay chains lead to
gamma-ray and neutrino emission. Isospin symmetry, which is in agreement with
Fermi´s original theory of pion production (where a thermal equilibrium of the
resulting pion cloud is assumed) relates the three multiplicities with an
equal sign, thus we simply write $\xi_{\pi}$.


We can go a step further and consider energy dependent multiplicities. In this
case, the relation between energy maxima is quite different from
Eq.(\ref{enumax}). According to Ginzburg \& Syrovatskii (1964)
\cite{ginzburg64}, for inelastic $pp$ interactions we can obtain the gamma ray
energy from the proton energy by \be E_p\approx 6 \;\xi_{\pi}(E_p)\
E_{\gamma}/K \label{ep-egamma}. \ee

The inelasticity coefficient is $K\sim0.5$ since on average a
leading nucleon and a pion cloud leave the interaction fireball each
carrying half of the total incident energy\foot{See ref.
\cite{gaisser90} for a discussion on this coefficient.}. For the
energy dependent pion multiplicity we will follow the prescription
adopted by Mannheim and Schlickeiser (1994) \cite{mann-schlick} \be
\xi_{\pi}\simeq (E_p/\rm GeV-1.22)^{1/4}.\label{multiplicity}\ee
Note however that for a proton energy of 1 PeV, the relation between
maxima given by eq. (\ref{ep-egamma}) would become $ E_p^{\rm
max}\sim 380 E_{\gamma}^{\rm max},$ and accordingly $ E_{\nu}^{\rm
max}\sim{1\over 760} E_p^{\rm max}$. This relation implies that the
1 TeV threshold detector IceCube should actually measure some
neutrinos provided the microquasar could accelerate protons up to
the PeV range. As a matter of fact, Eq. (\ref{multiplicity})
overestimates pion luminosity at energies higher than 10$^4$ GeV and
the 1/4 root should be relaxed for the proton energies involved in
our calculation \cite{mann-schlick}. Since we have no data about its
energy dependence we shall adopt a softer root growing law
\footnote{According to Begelman et al.(1990) \cite{begelman}, based
on Orth and Buffington (1976) \cite{orthbuff}, the multiplicity
should be kept below 15 for $E_p = 100$ TeV. In order to
analytically lessen Eq. (\ref{multiplicity}) to such values, we can
interpolate the 10 to 100 TeV range in a simple way with a 1/5
fractional power. This gives $\xi_{\pi}\approx 14$ for $E_p = 100$
TeV. For the next two decades, there are no confident approaches to
our knowledge so we extrapolate $\xi_{\pi}$ with a 1/6 root up to
the PeV maximal proton energies obtained in Eq.(\ref{epmax}). This
gives $\xi_{\pi} \approx$ 26 for $E_p = 50$ PeV (periastron). Notice
that a continuous expression for $\xi_{\pi}$ is needed in order to
evaluate both luminosity and neutrino signal from Eq.
(\ref{luminosity}) and Eq. (\ref{signal}). Notwithstanding, the
corresponding integrals are not really sensitive to this analytical
choice since the multiplicity effects are actually weak: on the one
hand, $\sigma_{pp}$ depends only logarithmically on $\xi_{\pi}$; on
the other, it modifies the upper limit of the integrals, but then
again as the integrand is quite steep the contribution of the queue
in a larger domain is below a few percent. Note finally that here we
conservatively adopted a $\log$ instead of a $\log^2\ $ $Ep$ -
dependence of the cross section.} for $\xi_{\pi}(E_p)$ which leads
to a more reliable relationship between neutrino and proton
energies. For the highest proton energies, as given by
Eq.(\ref{epmax}), one gets $E_{\nu}^{\rm max}\sim{1\over 156}
E_p^{\rm max}$ at periastron (see Fig.\ref{fig:BfieldEpmax}b). We
will promptly see that one can in fact expect for a significant
signal to noise relationship in less than one year of operational
time of a `km$^3$' telescope.


The point now is whether a microquasar might accelerate protons to such high
energies. As mentioned in Bednarek et al. (2005) \cite{bednaetal}, the maximum
energy at which a particle of charge $Z$ can be accelerated, can be derived
from the simple argument that the Larmor radius of the particle should be
smaller than the size of the acceleration region (Hillas 1984
\cite{hillas84}). If energy losses inside sources are neglected
\foot{Something that, given the size constraints and photon fields, is the
case in the jet of a source like LS I $+61$ 303.}, this maximum energy
$E^M_{15}$ (in units of $10^{15}$ eV) is related to the strength of the
magnetic field $B$ (in units of Gauss) and the size of the accelerating region
$R$ (in units of cm) by the following relationship
\begin{equation}
E^M_{15} \sim 0.3 \; 10^{-12} \beta~Z~B_{G}~R_{\rm cm} \label{hill}
\end{equation}
\noindent where $\beta$ is the velocity of the shock wave or the acceleration
mechanism efficiency. Hence, the maximum energy up to which particles can be
accelerated depends on the $B\times R$ product. Diffusive particle
acceleration through shocks may occur in many candidate sites with sizes
ranging from kilometers to megaparsecs. In the case of  LS I $+61 \;303$, we
adopt quasi-parallel shocks, $R\sim R_{\rm j}$ (jet radius), and $B$ given by
equipartition with the cold, confining plasma (see Bosch-Ramon et al. 2005b
\cite{br05b}).

For a complete calculation of the magnetic field, we will assume a
magneto-hydrodynamic mechanism for the ejection where both matter and field
follow adiabatic evolution when moving along the jet. In this conditions it is
reasonable to adopt equipartition between the magnetic field energy and the
kinetic energy of the jet. This leads directly to the following expression for
the magnetic field in the jet reference frame at different distances $z$ from
the compact object \cite{BK}\cite{br05b}

\be B(z)= \sqrt{8 \pi e_{\rm jet}},\label{B} \ee where \be e_{\rm jet}=
\frac{\dot{m}_{\rm jet}}{\pi R_{\rm j}^2(z) v_{\rm j} m_p} E_{\rm
k}\label{energydensity} \ee is the jet energy density for a cold proton
dominated jet. The mean cold proton kinetic energy, $E_{k}$, is taken to be
the classical kinetic energy of protons with velocity $v_{\rm j}= c \beta_{\rm
b}$. The ejected matter amounts to a fraction
of the ratio of accreted matter, namely $\dot{m}_{\rm jet}=q_{\rm j}
\dot{M}_{\rm c}$ with $q_{\rm j}\sim 0.1$.

Now, we can proceed and compute the maximum proton energy compatible with this
magnetic field by means of the gyro-radius identity\footnote{Notice that a
straightforward calculation of the proton losses shows that they do not
diminish the maximum energy since the size constraints are more important in
the present context.} \be E_p^{\rm max} = R(z)\ {\rm e} B(z).\label{epmax}\ee
This result will be used to compute the maximum jet gamma-ray and neutrino
energies which we will need in the next section to calculate both intensities.

In Figure \ref{fig:BfieldEpmax} we show the magnetic field (a) and maximum
proton energy (b) at the base of the jet in terms of the orbital position for
 LS I $+61$ 303.
\begin{figure*}[t]
\centering
\includegraphics[width=0.4\textwidth,height=3.5cm]{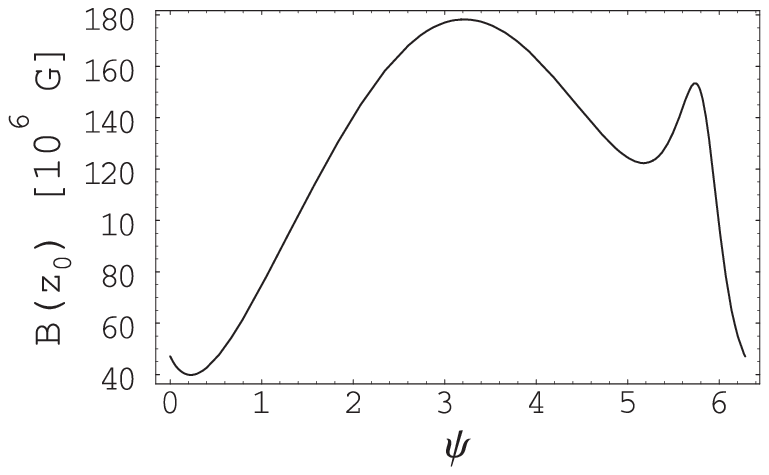}\hspace{2cm}
\includegraphics[width=0.4\textwidth,height=3.5cm]{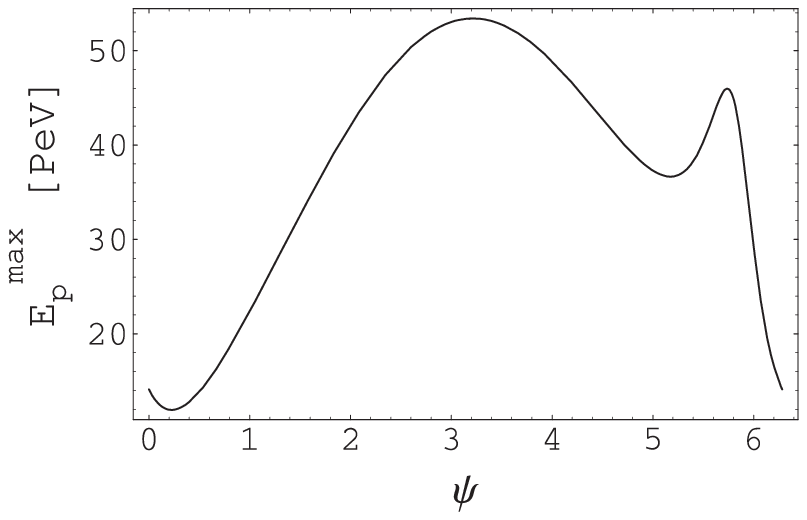}
\caption{Left (a): Maximal magnetic field in the jet, $B^{\rm max}(\psi)$, as
a function of the orbital position of the compact object (in Gauss). Right
(b): Highest proton energy compatible with the magnetic field (in PeV) as a
function of $\psi$.} \label{fig:BfieldEpmax}
\end{figure*}

As we can see in the figure, the maximum proton energy produced at the base of
injection amounts to more than 50 PeV at periastron and lies below 15 PeV near
apastron. Note that the minimum does not occur exactly at apastron due to the
magnetic field dependence. This is a consequence of the particular orbital
variation of the accretion rate. Furthermore, we see that there is a second
sharp maximum, readily identified with the function $1/v_{\rm rel}^3,$
 which grows dramatically  when the compact object gets far from the
primary star and moves more parallel to the weak stellar wind (see also
Mart{\'{\i}} \& Paredes 1995 \cite{paredes95} for a discussion of this
effect).

\section{Flux and event rate of neutrinos}

In order to obtain the spectral flux of neutrinos, we need an explicit
expression for the neutrino emissivity with the details therein. Instead, we
can take  eq. (\ref{conservation}) so as to extract an expression for
$I_{\nu}(E_{\nu})$. Let us write eqs. (\ref{conservation}) and (\ref{enumax}) as

\be L_\nu = D\ L_\gamma ,  \ee and

\be 2 E_{\nu} = E_{\gamma},\ee which result from the assumption that,
together, one ($\mu$) neutrino and one anti-neutrino carry half one
charged-pion energy (note that the number of ($\mu$) neutrinos resulting from
$\pi^+$ and $\pi^-$ decays is equal to the number of gamma rays coming from
$\pi^0$ since the number of pions produced is the same for the three flavors;
same holds for the flux of ($\mu$) anti-neutrinos).

From the last two equations we obtain \be I_{\nu}(E_{\nu})=4D\
I_{\gamma}(E_{\gamma}=2E_{\nu}). \ee For $pp$ interactions 1/3 of the proton
energy goes to each pion flavor and we have $D=1$. Thus we have for the
neutrino intensity \be I_{\nu}(E_{\nu},\theta)= 4 \int_{V} d V\
\frac{f_p\ \rho_w(r)}{m_p}\ q_\gamma(\vec{r}, 2E_{\nu},\theta) \ee

Now, we have to compute the convolution of the neutrino flux with the event
probability. For signal and noise above 1 TeV we obtain respectively

\be S =T_{\rm obs} A_{\rm eff} \frac 1{4\pi d^2}\ \frac
1{2\pi}\int_0^{2\pi}d\psi \int_{1 \rm TeV}^{E_{\nu}^{\rm max}(\psi)} dE_{\nu}\
I_{\nu}(E_{\nu},\psi)\ P_{\nu\rightarrow\mu}(E_{\nu}), \label{signal} \ee and
\begin{equation}
N=\bigg[T_{\rm obs} A_{\rm eff} \Delta\Omega \frac 1{2\pi}\int_0^{2\pi}d\psi
\int_{1 \rm TeV}^{E_{\nu}^{\rm max}(\psi)} dE_{\nu}\ F_B(E_{\nu})\
P_{\nu\rightarrow\mu}(E_{\nu})\bigg]^{1/2}, \label{noise}
\end{equation}
where $T_{\rm obs}$ is the observational time period, $A_{\rm eff}$ is the
effective area of the detector, $d$ is the distance to the system, and
$\Delta\Omega$ is the solid angle of the search bin. The function \be
F_B(E_{\nu})\leq 0.2\ (E_{\nu}/\rm GeV){}^{-3.21}\ \ {\rm GeV^{-1} cm^{-2}
s^{-1} sr^{-1}} \label{FB}\ee represents the $\nu_{\mu}+\bar\nu_{\mu}$
atmospheric flux (see Volkova 1980, and Lipari 1993 \cite{volkova, lipari}),
and \be P_{\nu\rightarrow\mu}(E_{\nu})= 1.3\ 10^{-6}\ (E_{\nu}/\rm
TeV){}^{0.8} \ee is the probability that a neutrino of energy $E_{\nu}\sim 1 -
10^3$ TeV, on a trajectory through the detector produces a muon (see Gaisser
et al. 1995 \cite{Physrep}).

As we have already discussed, the value of the maximal neutrino energy is a
function of orbital position of the jet and is given by \be E_{\nu}^{\rm
max}(\psi)\approx \frac{K}{12\xi_{\pi}(\psi)}\ E_p^{\rm max}(\psi).\ee Our
estimate for this expression is $E_{\nu}^{\rm max}\approx 85$ TeV at
periastron, and about 20 TeV at apastron (see Fig.\ref{fig:enumax}).

\begin{figure*}[t]
\centering
\includegraphics[width=0.4\textwidth,height=3.5cm]{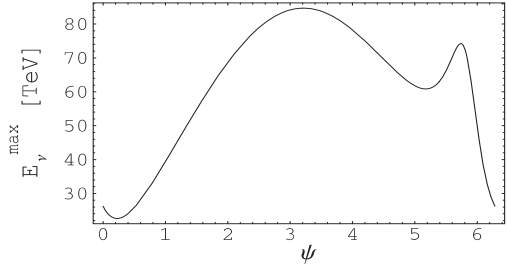}
\caption{Maximal neutrino energies $E_{\nu}^{\rm max}$ along the orbit (in TeV).}
\label{fig:enumax}
\end{figure*}

If we now  set the features of a km-scale detector such as IceCube
\cite{icecube} in the model ($A_{\rm eff}= 1$ km$^2$ and $\Delta\Omega\approx
3\ 10^{-4}$sr), and the distance to LS I $+61$ 303 ($d=2$ kpc) we get \be
S/N\mid_{1 \rm year}\geq 9.42 \ee for the signal to noise ratio in a one-year
period of observation. This is consistent with the upper limit recently
reported by Ackermann et al. (2005) from AMANDA-II experiment \cite{amanda2}.

Note that the present prediction is restricted to $\mu$-neutrino production at
the source. As a matter of fact, there is nowadays strong experimental
evidence of the existence of neutrino oscillations which occur if neutrinos
are massive and mixed. In 1998 (see Super-Kamiokande \cite{SK98}) and in 2002
(see K2K \cite{K2K}) it has been observed, respectively, the disappearance of
atmospheric and laboratory muon-neutrinos as expected from flavor
oscillations. Also in 2002, the SNO experiment \cite{SNO} provided solid
evidence of solar electron-neutrino oscillations to other flavors. Solar and
atmospheric neutrino flux suppression can be explained in the minimal
framework of three-neutrino mixing in which the active flavor neutrinos $\nu_e
, \nu_{\mu}$ and $ \nu_{\tau}\ $ are unitary linear combinations of three mass
(Majorana) eigenstates of the neutrino lagrangian \cite{mixing}.

The expected ratios at sources of high-energy neutrino fluxes from $pp$
collisions are $1 : 2 : 10^{-5}\ $ for the $e, \mu, \tau\ $ neutrino flavors,
in the range 1 GeV $\leq E_{\nu} \leq 10 ^{12} $ GeV. The neutrino oscillation
effects imply that one should measure different values depending on the
distance to the astrophysical source. The estimate is that these ratios become
in average $1 : 1 : 1\ $  for $L(\rm pc)/E(\rm GeV)\geq 10^{-10},\ $ where the
distance is in parsecs (see Athar et al. 2005 \cite{athar05}). Also recently,
a detailed analysis of supernova remnants reported by Costantini and Vissani
(2005) \cite{cv05} claims a 50$\%$ of muon-neutrino plus muon anti-neutrino flux
reduction due to flavor oscillations along astrophysical distances. Note
finally, that the neutrino signal measured at Earth might be further
attenuated due to matter absorption \foot{Separate (in matter) flavor
oscillations effects for its path across the Earth should be also considered
for an exhaustive analysis.} (see both references above)\foot{According to the
central values of the mixing angles reported in Costantini and Vissani 2005
\cite{cv05} it is not clear if there could also be a slight contribution to
muon-neutrino detection due to electron-neutrino oscillation. As for
tau-neutrinos, they should be already highly suppressed at the source.}.


\section{Luminosity and opacity}

As a result of the discussion above, our predictions for both neutrino and
gamma ray luminosities between $E_{\nu,~\gamma}^{\rm min}= 1, 2$ TeV and
$E_{\nu,~\gamma}^{\rm max}$ amount to 5 10$^{34}$ erg s$^{-1}$ for a source
such as LS I $+61$ 303.

 The total neutrino luminosity is related to the spectral
neutrino {\it intensity}  by means of
 \be L_\nu =\int_{{E_{\nu}^{\rm min}}}^{E_{\nu}^{\rm max}} dE_{\nu}\ E_{\nu}\
 I_{\nu}(E_{\nu})\ee as we mentioned in Sect. \ref{model}. Since we do not have
 an expression for $I_{\nu}(E_{\nu})$, we can use eq. (\ref{conservation})
and relate it to the gamma ray luminosity
 \be L_\gamma =\int_{{E_{\gamma}^{\rm min}}}^{E_{\gamma}^{\rm max}} dE_{\gamma}\
E_{\gamma}\ I_{\gamma}(E_{\gamma})\ee which we know in more detail. Now,
 there is in fact an experimental constraint on this quantity, as
 recently analyzed by Fegan et al. (2005) \cite{fegan05}. There, it is claimed that,
 above 0.35 TeV, the total flux of
gamma rays that can be produced must satisfy \be \frac{1}{4\pi
d^2}\int_{{0.35\rm TeV}}^{\infty} dE_{\gamma}\ I_{\gamma}(E_{\gamma})\leq 1.7\
10^{-11}{\rm\ ph\ cm^{-2} s^{-1}}.\ee Our calculations for gamma rays give
about four times this value, so we need to show that most of the high-energy
gamma rays get absorbed. Of course, the absorption mechanism should not modify
the flux of neutrinos. Infrared photon fields can be responsible for TeV
photon absorption in the source. To show it, we calculate the optical depth
$\tau$ within the circumstellar disk for a photon with energy $E_{\gamma}$

\begin{equation}
\tau(\rho,\,E_{\gamma})=\int_{E_{\rm min}(E_{\gamma})}^\infty dE_{\rm
ph}\,\int_{\rho}^\infty d\rho\ n_{\rm ph}(E_{\rm ph},\rho)\
\sigma_{e^-e^+}(E_{\rm ph},E_{\gamma}), \label{tau}
\end{equation}
where $E_{\rm ph}$ is the energy of the ambient photons, $n_{\rm ph}(E_{\rm
ph},\rho)$ is their density at a distance $\rho$ from the neutron star, and
$\sigma_{e^-e^+}(E_{\rm ph},E_{\gamma})$ is the photon-photon pair creation
cross section given by
\begin{equation}
\sigma_{e^+e^-}(E_{\rm ph}, \;E_{\gamma})=\frac{\pi
r_0^2}{2}(1-\xi^2)\left[2\xi(\xi^2-
2)+(3-\xi^4)\ln\left(\frac{1+\xi}{1-\xi}\right)\label{sigmaee} \right],
\end{equation}
where $r_0$ is the classical radius of the electron and
\begin{equation}
\xi=\left[1-\frac{(m_e c^2)^2}{E_{\rm ph} E_{\gamma}}\right]^{1/2}.
\end{equation}
In Eq. (\ref{tau}), $E_{\rm min}=m_e^2 c^4/E_{\gamma}$ is the threshold energy
for pair creation in the ambient photon field. This field can be considered as
formed by two components, one from the Be star and the other from the cold
circumstellar disk\foot{There is, in fact, a third component corresponding to
the emission from the heated matter (hot accreting matter impacting onto the
neutron star) which can be approximated by a Bremsstrahlung spectrum
\begin{equation}
n_{\rm ph,3}(E_{\rm ph},\rho)=\frac{L_X\,E_{\rm ph}^{-2}}{4\pi
c\,\rho^2\,e^{E_{\rm ph}/E_{\rm cut-off}}}\mbox{  for $E_{\rm ph}\geq 1$ keV},
\end{equation}
where $L_X$ is the total luminosity. This component does not significantly
affect the propagation of TeV gamma-rays so we neglect this contribution.},
$n_{\rm ph}=n_{\rm ph,1}+ n_{\rm ph,2}$. Here,
\begin{equation}
n_{\rm ph,1}(E_{\rm ph},\rho)= \left(\frac{\pi B(E_{\rm ph})}{hc\,E_{\rm
ph}}\right)\frac{R_\star ^2}{\rho^2+r^2-2\rho r \sin\theta}\;,
\end{equation}
is the black body emission from the star, and
\begin{equation}
n_{\rm ph,2}(E_{\rm ph},\rho)= \left(\frac{\pi B(E_{\rm ph})}{hc\,E_{\rm
ph}}\right)\frac{r^2}{\rho^2}\;, \label{ndisk}
\end{equation}
corresponds to the emission of the circumstellar disk. In both cases we adopt
\begin{equation}
B(E_{\rm ph})= \frac{2 E_{\rm ph}^3}{(hc)^2\,(e^{E_{\rm ph}/kT_{\rm eff}}-1)}
\end{equation}
where $T_{\rm eff,1}=22500$ K and $T_{\rm eff,2}=17500$ K (Mart\'{\i} \&
Paredes 1995 \cite{paredes95}).

\begin{figure*}[t]
\centering
\includegraphics[width=0.4\textwidth,height=3.5cm]{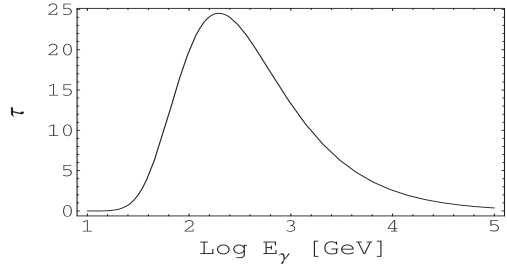}
\caption{Optical depth at periastron, as a function of $\log(E_{\gamma}/\rm
GeV)$.} \label{fig:tau}
\end{figure*}

The mentioned value for $T_{\rm eff,2}$ is valid in the inner region of the
disk and thus reliable only near periastron (as one gets far from the central
star the temperature of the disk gets lower). Hence we adopt $r=r_{\rm
periastr}$. In any case, computing the opacity at periastron, where the
luminosity is maximal, will be enough for our purposes of proving that the
total gamma-ray luminosity predicted in our model is below the Fegan et al.'s
constraint \cite{fegan05}.  Figure \ref{fig:tau} shows the $E_\gamma$
dependence of the optical depth  at periastron for an observer at $\theta_{\rm
obs}=30^\circ$ with respect to the jet axis. The optical depth remains above
unity for a wide range of photon energies but a sharp knee takes place at
$E_\gamma\approx 250$ GeV. For TeV gamma energies it still has an important
effect on the luminosity. In particular, the opacity-corrected total flux
above $E_\gamma=$ 350 GeV drops from 7.16 $10^{-11}$ to 1.42 $10^{-12}$ ph
cm$^{-2}$ s$^{-1}$, lying well below the Fegan et al.'s threshold.

Other observing windows, however, could reveal the presence of a hadronic
gamma-ray source at the position of LS I $+61 \, 303$. In particular, between
1 and 50 GeV the opacity is sufficiently low as to allow a relatively easy
detection by instruments like the ground-based MAGIC telescope and the LAT
instrument of GLAST satellite. The source is in the Northern Hemisphere, hence
out of the reach of HESS telescopes, which recently detected the microquasar
LS 5039 \footnote{Late after the completion of this paper we became aware of a
subsequent upload \cite{demente} where the potential TeV neutrino source LS
5039 is considered.}. LS I $+61 \, 303$ is an outstanding candidate to
corroborate that high-energy emission is a common property of microquasars.
Its location, in addition, makes of it an ideal candidate for IceCube. A
neutrino detection from this source would be a major achievement, which would
finally solve the old question on whether relativistic protons are part of the
matter content of the energetic outflows presented by accreting compact
objects.

\section{Conclusions}

We have analyzed the possible origin of high-energy neutrinos and gamma rays
coming from a galactic source. Our specific subject for the analysis has been
the microquasar LS I $+61 \;303$ because it is a well-studied object of unique
characteristics and an appropriate candidate to make a reliable prediction of
the neutrino flux from a galactic point source. We performed our calculations
within a purely hadronic framework and showed how neutrino observatories like
IceCube can establish whether TeV-gamma rays emitted by microquasars are the
decay products of neutral pions. Such pions are produced in hadronic jet -
wind particle interactions. We improved previous predictions by considering
realistic values for the parameters of the system and energy dependent pion
multiplicities particularly significant at high energies. Above 1 TeV, we
obtained a mean-orbital $\nu$-luminosity of 5 $10^{34}$ erg/s which can be
related to an event rate of 4--5 muon-type neutrinos per kilometer-squared per
year if we take into account neutrino oscillations. The upper limit of
integration depends on the orbital position and is a function of the largest
magnetic gyro-radius compatible with the jet dimensions. As a consequence, the
maximal neutrino energies here considered range between 20 and 85 TeV along
the orbit. Opacity effects on the associated gamma radiation are due to the
infrared photon field at the source and result in a significant attenuation of
the original $\gamma$-ray signal. Nonetheless, LS I $+61 \, 303$ might be
detectable at low GeV energies with instruments like MAGIC and GLAST. Such a
detection, would be crucial to test current ideas of particle acceleration in
compact objects.


\subsection*{Acknowledgements}
H.R.C. thanks financial support from FUNCAP and CNPq, Brazil.
G.E.R. and M.O. are supported by CONICET (PIP 5375) and ANPCyT (PICT 03-13291), Argentina.

\clearpage
\begin{table} 
\begin{center}
\caption{Basic parameters assumed for the model}
\begin{tabular}{lll}
\hline

\hline
Parameter & Symbol  & Value  \\
\hline
Mass of the compact object & $M_{\rm c}$ & 1.4 $M_{\odot}$\\
Jet's injection point $^{\,1}$& $z_0$ & 50 $R_{\rm g}$  \\
Initial radius & $R_0$ & $z_0/10$ \\
Mass of the companion star & $M_\star$ & 10 $M_{\odot}$\\
Radius of the companion star & $R_\star$ &10 $R_{\odot}$ \\
Effective temperature of the star & $T_{\rm eff}$& 22500 K\\
Density of the wind at the base& $\rho_0$& $10^{-11}$ gr cm$^{-3}$\\
Initial wind velocity$^2$ & $v_0$ & 5 km s$^{-1}$\\
Jet's Lorentz factor & $\Gamma$ & 1.25 \\
Minimum proton energy & ${E'}_p^{\rm min}$ & 1 GeV \\
Penetration factor & $f_{\rm p}$ & 0.1 \\
Orbital axis (cube) & $a^3$ & $\frac{P^2  G (M_\star + M_{\rm c})}{4 \pi^2}$\\
Eccentricity & $e$ & 0.72 \\
Orbital period & $P$ & 26.496 d\\
Relativistic jet power coupling & $q_j$ & 0.01\\
Index of the jet proton distribution &$\alpha$& 2.2\\
\hline \multicolumn{3}{l} {$^1\ \ R_{\rm g}=GM_{\rm c}/c^2$.}\cr
\multicolumn{3}{l} {$^2$ The wind velocity increases from this value up to the escape}\cr
\multicolumn{3}{l} { velocity according to the velocity law given in the text.}\cr
\end{tabular}
\end{center}
\label{t1}
\end{table}

\end{document}